\documentstyle[12pt]{article}       
       
\newlength{\dinwidth}       
\newlength{\dinmargin}       
\def\lapproxeq{\lower .7ex\hbox{$\;\stackrel{\textstyle       
<}{\sim}\;$}}       
\def\gapproxeq{\lower .7ex\hbox{$\;\stackrel{\textstyle       
>}{\sim}\;$}}       
\def\be{\begin{equation}}       
\def\ee{\end{equation}}       
\def\bea{\begin{eqnarray}}       
\def\eea{\end{eqnarray}}       
\def\ktbold{\mbox{\boldmath${k}$}_T}       
       
\begin{document}       
\titlepage       
\begin{flushright}       
DTP/98/04 \\       
February 1998 \\       
Revised March 1998 \\  
\end{flushright}       
       
\vspace*{2cm}       
       
\begin{center}       
{\Large \bf Higher twists in deep inelastic scattering}       
       
\vspace*{1cm}       
A.D.Martin$^a$ and M.G.Ryskin$^b$\\       
       
\vspace*{0.5cm}       
$^a$ Department of Physics, University of Durham, Durham,    
DH1 3LE\\       
$^b$ Petersburg Nuclear Physics Institute, Gatchina, St.    
Petersburg, 188350,        
Russia       
\end{center}       
       
\vspace*{2cm}       
       
\begin{abstract}       
We perform an exploratory study of higher twist    
contributions       
to deep inelastic scattering.  We estimate the size of    
two       
major sources of higher twist, namely absorptive    
corrections       
and the vector meson dominance (VMD) contribution.  We    
find       
that they give a sizeable higher twist component of    
$F_2$.  For example at $x = 0.01$      
it is about 8\% (17\%) at $Q^2 = 10$~GeV$^2$ (4~GeV$^2$),    
reaching up to 27\% at      
$x = 10^{-4}$ and $Q^2 = 4$~GeV$^2$.  At the smaller $x$    
value the largest      
contribution comes from absorptive corrections, while at    
the larger values of $x$ the      
VMD term dominates.      
       
\end{abstract}       
       
\newpage       
       
At large $Q^2$ the cross section for deep inelastic    
scattering (DIS) is to a        
good approximation described by just the twist-2    
component of the structure        
function. That is\footnote {Since we are interested in    
the possible higher twist       
effects on parton analyses we concentrate on the cross    
section for the        
absorption of transversely polarised photons,    
$\sigma_T$.}      
\begin{equation}       
\sigma_T(\gamma^*p) \; = \; \frac{4\pi^2\alpha}{Q^2}    
F_T(x,Q^2)       
\label{eq:a1}     
\end{equation}       
with $F_T \simeq F_T^{(2)}$, where $\alpha$ is the    
electromagnetic        
coupling. Indeed in extracting parton distributions from    
DIS data it is        
commonly assumed that (\ref{eq:a1}) is exact and that    
$F_T$ is given entirely by        
twist-2. To leading order in QCD we have       
\begin{equation}       
F_T^{(2)} = \sum_{q} e_q^2 x[q(x,Q^2)+\bar{q}(x,Q^2)]       
\label{eq:a2}     
\end{equation}       
where $q$ and $\bar{q}$ are the quark and antiquark        
distributions and $ee_q$ is the charge of the quark. For    
sufficiently        
small values of $Q^2$ the higher twist (4,6,...)    
components of $F_T$,       
defined by       
\begin{equation}       
F_T = F_T^{(2)} + \frac{F_T^{(4)}}{Q^2} +    
\frac{F_T^{(6)}}{Q^4} + ...,       
\label{eq:a3}     
\end{equation}       
would be expected to give noticeable contributions to    
$\sigma_T$.       
Surprisingly, even though the DIS data have become much    
more precise, the recent       
global analyses still show no necessity for higher twist    
contributions ---       
despite including data at remarkably low values of $Q^2$.    
Moreover parametric        
fits of $F_2$ data \cite{NMC} have found very small    
values of $F_2^{(4)}$       
at low $x$. Our objective is to explore the role, and to    
estimate the size, of       
the higher twist terms.       
       
>From the point of view of the Wilson Operator Product    
Expansion (OPE) the higher       
twist terms correspond to operators describing a larger    
number of partons. Say,        
for twist-4 we must consider operators with four quarks    
$\left<N|q\bar{q}q       
\bar{q}|N\right>$ or four gluons $\left<N|gggg|N\right>$,    
and so on.       
Strictly speaking for each new twist and new operator we    
have to specify a new        
input function which should be determined by a global fit    
to the DIS data.        
Unfortunately the data are not yet precise enough to    
determine more than the        
leading twist, $F_T^{(2)}$, decomposition in terms of    
partons.       
       
Rather we will discuss two effects which are expected to    
give the dominant        
twist-4 contributions in the HERA domain. The    
contributions arise from (a)        
absorption corrections and (b) the vector meson dominance    
(VMD) contribution.        
We estimate their size below, but first we show their    
$1/Q^4$ behaviour.        
       
At low $x$ the behaviour of $\sigma_T(\gamma^*p)$ is    
controlled by the evolution       
of gluons in the $t$ channel. The evolution equation may    
be written       
\begin{eqnarray}       
xg(x,Q^2) = xg(x,Q^2_0) + \int^1_x \frac{dz}{z}    
\int^{Q^2}_{Q^2_0}        
\frac{dQ^{\prime2}}{Q^{\prime2}} \    
\frac{\alpha_S(Q^{\prime2})}{2\pi} \        
\frac{x}{z} \ g(\frac{x}{z},Q^{\prime2})\ P_{gg}(z)    
\nonumber \\       
-A \int^1_x \frac{dx^\prime}{x^\prime} \int^{Q^2}_{Q_0^2}    
\frac{dQ^{\prime2}}       
{Q^{\prime4}} \alpha^2_S(Q^{\prime2})[x^\prime    
g(x^\prime,Q^{\prime2})]^2       
\label{eq:a4}     
\end{eqnarray}       
where the first two terms correspond to the conventional    
twist-2        
DGLAP evolution with gluon-gluon splitting function    
$P_{gg}(z)$, while the last       
term with the negative sign takes into account the    
higher-twist absorptive        
corrections\footnote{Such a form was first introduced in    
Ref.~\cite{GLR}, and       
studied further in Ref.~\cite{MQ}.}.  The value of $A$ is    
discussed below. The last      
term corresponds to gluon recombination and is shown    
schematically in Fig.~1, where      
the four gluon twist-4 structure is evident. The extra    
$Q^{\prime2}$ in the      
denominator of the absorptive term reflects the small    
probability to find an additional      
gluon in a small domain of transverse size $1/Q^\prime$.     
Strictly speaking the last      
term contains both twist-2 and twist-4 components coming    
from evaluating the      
integral at the lower limit $Q_0^2$ and the upper limit    
$Q^2$ respectively.  The      
twist-2 component may be regarded as an extra input for    
the twist-2 DGLAP      
evolution.  On the other hand the contribution from the    
upper limit results from the      
evolution of the four gluon state from $Q_0^2$ to $Q^2$,    
and is therefore manifestly      
twist-4.     
     
To identify the twist-2 and twist-4 parts of the    
absorptive correction we simply rewrite      
the last term in (\ref{eq:a4}) in the form     
\be     
-A \int_{Q_0^2}^{Q^2} \cdots \; = \; -A    
\int_{Q_0^2}^{\infty} \cdots \: + \: A      
\int_{Q^2}^{\infty} \cdots ~.     
\label{eq:b4}     
\ee     
The negative first term combines with the input    
distribution, $xg (x, Q_0^2)$, in      
(\ref{eq:a4}) to give a new initial condition for the    
twist-2 evolution, while the      
positive second term is the twist-4 component.  Thus    
although the whole absorptive      
correction is negative, the twist-4 component is itself    
positive.     
       
A second well-known higher twist contribution to    
$\sigma_T(\gamma^*p)$        
comes from the Vector Meson Dominance (VMD) term       
\begin{equation}       
\sigma_T({\rm VMD}) = \pi\alpha \sum_V    
\frac{M_V^4\sigma_V(s)}       
{\gamma_V^2(Q^2+M_V^2)^2}       
\label{eq:a5}     
\end{equation}       
where the sum is over vector mesons $V$ of mass $M_V$ and where $\gamma_V$   
specifies the $\gamma-V$ coupling. $\sigma_V(s)$ is the total $Vp$ cross section at   
centre-of-mass energy $\sqrt{s}$. The $\sigma_T({\rm VMD})$ component   
dominates $\sigma_T(\gamma^*p)$ at very low $Q^2$.  The space-time picture of the   
VMD contribution is compared with that of the conventional (twist-2) DGLAP   
contribution in Fig.~2. For the DGLAP contribution of diagram (a)  the upper quark   
propagator $1/(Q + k)^2 \simeq 1/Q^2$, so that the distance between points 1 and 2,   
$\Delta r_{12} \sim 1/Q$, is very small. Moreover in terms of old-fashioned   
perturbation theory, the invariant mass of the produced $q\bar{q}$ system $M \sim   
Q$. In contrast in the small region of phase space corresponding to $M^2 \sim M_V^2   
\ll Q^2$ the left-hand diagram of 2(a) may be drawn as in the left-hand diagram of   
2(b), where the space-time development of the VMD contribution is evident.  The   
photon first creates a  light $q\bar{q}$ pair long before the interaction with the target.   
The presence of the four quark twist-4 structure is evident in the schematic right-hand   
diagram of 2(b), and is reflected by the $1/Q^4$ behaviour of (\ref{eq:a5}) at large   
$Q^2$.  In this case, which corresponds to large distances and rather small transverse   
momentum $k_T$ of the quarks, it is better to deal with constituent quarks with   
masses $m_{u,d} =  350$~MeV and $m_s = 500$~MeV. \\     
       
\noindent {\large \bf Absorptive or gluon rescattering    
effects}       
       
In general the four $t$ channel gluons shown in the lower    
part of Fig.~1 interact       
with each other. It is convenient to re-organize the    
perturbative expansion and       
to consider first the interaction between the gluons in    
each pair separately        
and then to consider the interactions between the two    
gluon ladders. The sign       
and the value of each contribution depends on the colour    
structure. Each pair       
of gluons may form a singlet  (\ref{eq:a1}), a symmetric    
or antisymmetric octet        
($8_s,8_a$), decuplets ($10,\overline{10}$) or a 27 colour    
state.  The various       
colour configurations have different energy dependences       
\begin{equation}       
\sigma \sim(s^\prime)^{2\alpha(0)-2}       
\label{eq:a6}     
\end{equation}       
depending on the intercept $\alpha(0)$, where    
$\sqrt{s^\prime}$ is        
the energy shown in Fig.~1.  Information on the    
intercepts comes from the BFKL      
equation.  Now the BFKL kernel $K$ contains two parts ---    
the virtual (one-loop)      
correction, which results in the reggeization of the $t$    
channel gluons, and a term      
describing real $s$ channel gluon emission.  In general    
the intercepts may be written in      
the form     
\be     
\alpha_i (0) \; = \; 1 \: + \: \alpha_S \left [ c_i    
\langle K_{\rm real} \rangle \: - \: \langle      
K_{\rm virtual} \rangle \right ]     
\label{eq:b6}     
\ee     
where the kernels are averaged over the corresponding    
BFKL eigenfunctions.  The      
virtual correction is negative and does not depend on the    
total colour charge of the      
gluon pair, while the colour factor $c_i$ for real    
emission is equal to 3, $\left (      
\frac{3}{2}, \frac{3}{2} \right )$, 0 and --1 for the    
singlet, octets ($8_a, 8_s$),      
decuplets and 27-plet configurations respectively.  Due    
to the Regge      
bootstrap\footnote{In other words the octet BFKL    
amplitude self-consistently      
reproduces the original gluon trajectory with $\alpha_8    
(0) = 1$.} property of the     
BFKL      
equation, the octet intercepts are $\alpha_8 (0) = 1$ ---    
the virtual correction cancels      
the real emission part of the kernel.  On the other hand    
from (\ref{eq:b6}) we see that      
the intercepts of the 10 and 27 configurations are less than    
1 and their        
contributions therefore decrease with energy. Only the    
singlet-singlet        
configuration gives an amplitude which grows faster than    
the twist-2        
contribution as $x \rightarrow 0$, since the singlet    
intercept        
$\alpha_1(0) > 1$.       
     
To estimate the size of the twist-4 term in the gluon    
evolution equation (\ref{eq:a4})      
we therefore need to evaluate the factor $A$ in the last    
term for the singlet-singlet      
configuration.  First, it contains a colour factor of    
9/16 corresponding to the coupling      
of the four gluon state to the two $t$ channel gluons.     
The best way to obtain this      
factor is to consider the cross section for heavy photon    
dissociation \cite{RB,W} and      
to use the AGK cutting rules \cite{AGK}, which have been    
justified in QCD in refs.       
\cite{BW,BR}.  To explain the remaining content of $A$ it    
is convenient to write     
\begin{equation}       
A = \frac{9}{16} \ \frac{CK}{B}.       
\label{eq:a7}     
\end{equation}       
The dimensional factor $B$ (which compensates for the    
extra $Q^{\prime2}$ in       
the denominator in the last term of (\ref{eq:a4})) comes from the integral over the    
momentum $t = (p-p^\prime)^2$       
transferred through the \lq\lq pomeron" loop (which is    
indicated by the       
circular arrow on Fig.~1). In accordance with the    
measurements of heavy photon       
dissociation at HERA \cite{HERA} we use       
\begin{equation}       
B \; = \; \frac{d\sigma(0)}{dt} \: {\Bigg /} \: \int    
\frac{d\sigma(t)}{dt} \: dt     
\; \simeq \; 7.2~{\rm GeV}^{-2}.       
\label{eq:a8}     
\end{equation}       
Until now we have considered only \lq\lq elastic"    
$\gamma^*p       
\rightarrow Xp$ proton interactions. However the    
probability of dissociation        
of the target proton is not negligible. It is of order    
50-70\% of the        
\lq\lq elastic" interaction \cite{PDISS}. Thus we include    
a factor $C$ = 1.6 in       
(\ref{eq:a7}). Finally we include in the coefficient $A$    
the effect of pomeron-     
pomeron rescattering (shown in Fig.~3). Such rescattering    
partially fills the rapidity       
gap and so this contribution is not usually included in    
the diffractive        
component $F_2^D$ of $F_2$. The contribution increases    
with energy, but        
fortunately the dependence is rather slow ($\sim    
1+0.45({\rm ln}1/x)^{1/4}$) and        
so we may include it as a constant $K$ factor. In the    
HERA domain $K = 1.63$ to      
1.76 \cite{KFAC} and so we let $K=1.7$ in (\ref{eq:a7}).    
Interestingly our resultant      
value of $A$ is in close agreement with the value    
($81/16R^2 \simeq 0.2~{\rm      
GeV}^{-2}$ for $R=1$ fm) used in phenomenological    
analyses of absorptive      
corrections \cite{KMRS}.       
       
To compute the higher twist $\Delta F_T$ contribution to    
$F_T$ we start with       
an ordinary (twist-2) DGLAP fit to DIS and treat the last    
term of the       
evolution equation (\ref{eq:a4}, \ref{eq:b4}) as a small    
correction       
\begin{equation}       
x \Delta g(x,Q^2) =  A \int^1_x    
\frac{dx^\prime}{x^\prime} \int^{\infty}_{Q^2}       
\frac{dQ^{\prime2}}{Q^{\prime4}}    
[\alpha_S(Q^{\prime2})x^\prime g(x^\prime,       
Q^{\prime2})]^2.       
\label{eq:a9}     
\end{equation}       
Now at small $x$ the quark density is dominantly driven    
by the        
$g \rightarrow q\bar{q}$ transition. Though a full    
calculation is        
possible, for our exploratory study       
it is sufficient to use the approximation due to Prytz    
\cite{PRY},       
that is we evaluate       
\begin{equation}       
\frac{\Delta F_T(x,Q^2)}{F_T(x,Q^2)} \ \simeq \    
\frac{\Delta g(2x,Q^2)}       
{g(2x,Q^2).}       
\label{eq:a10}     
\end{equation}       
Our pure gluonic estimate of $F_2^{(4)}$ omits the two    
jet dissociation,        
$\gamma \rightarrow q \bar{q}$ , contribution    
corresponding to Fig.~4.       
This graph contributes to the diffractive structure    
function $F_2^D$ but not        
to the gluon distribution. Rather it contributes to the    
quark distribution,       
but with a strength which is supressed relative to the    
corresponding gluon       
distribution by the colour factor $(C_F/C_A)^2 \sim 1/5$.    
Moreover it gives        
a higher twist contribution to both $F_L$ and $F_T$. The    
explicit form of this       
small correction to $F^{(4)}$ has been evaluated in refs.    
\cite{FLFT,LMRT}, and       
should be included in a detailed calculation of the    
higher twist contribution. \\      
       
\noindent {\large \bf Vector Meson Dominance    
contribution}       
       
In the vector meson dominance contribution (\ref{eq:a5})    
to higher twist, usually only      
the       
$\rho,\omega$ and $\phi$ mesons are included       
\begin{equation}       
\sigma_T(\gamma^*p) = \pi \alpha    
\sum_{V=\rho,\omega,\phi} \ \frac{M_V^4        
\sigma_V(s)}{\gamma_V^2(Q^2+M_V^2)^2}.     
\label{eq:a11}       
\end{equation}       
We evaluate the $\rho p, \omega p$ and $\phi p$ cross    
sections       
$(\sigma_V)$ at centre-of-mass energy $\sqrt{s}$ as    
described in the footnote to      
eq.~(\ref{eq:a18}). Of course there will be some    
contribution from the higher mass      
vector meson resonances but there are expectations that    
they will be suppressed by       
smaller values of $1/\gamma_V^2$ and/or possibly smaller    
$Vp$ cross sections       
$\sigma_V$.       
       
We can also make an alternative estimate of this non-   
perturbative higher twist       
contribution based on hadron-parton duality.  Since for small $x$ the    
$\gamma^* \rightarrow      
q\bar{q}$ fluctuations occur over a much longer time    
scale than the interaction of the      
$q\bar{q}$ pair with the target proton, we may use    
hadron-parton duality to write the      
$\gamma^* p$ cross section in terms of a dispersion    
relation with respect to the      
invariant $q\bar{q}$ mass $M$ \cite{BB},       
\begin{equation}       
\sigma(\gamma^*p) = \sum_q \int^{\infty}_0    
\frac{dM^2}{(Q^2+M^2)^2}        
\ \rho(s,M^2) \sigma_{q \bar{q}+p}(s,M^2),       
\label{eq:a12}     
\end{equation}       
where $\sigma_{q \bar{q}+p}$ is the cross section for the        
scattering of the $q\bar{q}$ system on the proton and    
where the spectral function        
$\rho$ represents the density of $q\bar{q}$ states.  We    
may use perturbative QCD to      
evaluate (\ref{eq:a12}).  The cross section is given by    
the probability $|{\mathcal      
M}|^2$ of the $\gamma^* \rightarrow q\bar{q}$ transition multiplied by the   
imaginary part of the forward amplitude describing the $q\bar{q}$-proton    
interaction     
\be     
A_{q\bar{q} + p} \; = \; i s \sigma_{q\bar{q} + p}.     
\label{eq:b12}     
\ee     
For transversely polarized photons the amplitude of the    
$\gamma^* \rightarrow      
q\bar{q}$ transition is     
\begin{equation}       
{\mathcal M}_T = \frac{\sqrt{z(1-z)}}{\bar{Q}^2 + k^2_T}    
\        
\bar{u}_{\lambda}(\gamma .    
\epsilon_{\pm})u_{\lambda^\prime}       
= \frac{(\epsilon_{\pm}.k_T)[(1-2z)\lambda \pm 1]        
\delta_{\lambda,-\lambda^\prime}       
+ \lambda m_q \delta_{\lambda\lambda^\prime}}        
{\bar{Q}^2 + k_T^2},     
\label{eq:a13}       
\end{equation}       
where the $q$ and $\bar{q}$ longitudinal momentum    
fractions and        
transverse momenta are $z,\ktbold$ and $(1-z), -\ktbold$.       
We use the notation of Ref.~\cite{LMRT}, which was based    
on the earlier work of       
Ref.~\cite{BL}. Namely $\bar{Q}^2$ and the photon    
polarization vectors are        
given by       
\begin{eqnarray}       
\label{eq:a14}     
\bar{Q}^2 = z(1-z)Q^2+m^2_q  \\       
\epsilon_T = \epsilon_{\pm} = \frac{1}{\sqrt{2}}    
(0,0,1,\pm i),       
\label{eq:a15}     
\end{eqnarray}       
and where $\lambda, \lambda^{\prime} = \pm 1$ according    
to whether       
the $q, \bar{q}$ helicities are $\pm \frac{1}{2}$.       
       
In terms of the quark momentum variables we thus obtain     
\begin{equation}       
\sigma_T(\gamma^*p) = \sum_q \alpha \frac{e^2_q}{2\pi}    
\int dz\ dk^2_T \        
\frac{[z^2 + (1-z)^2]k^2_T+m^2_q}{(\bar{Q}^2 + k^2_T)^2}    
\        
N_c \sigma_{q\bar{q}+p} (s, k^2_T)       
\label{eq:a16}     
\end{equation}       
where the number of colours $N_c = 3$.  Before evaluating    
(\ref{eq:a16}) let us relate      
this expression to the dispersion relation form of    
(\ref{eq:a12}).  We may use     
\begin{equation}       
M^2 = \frac{k^2_T + m^2_q}{z(1-z)}     
\label{eq:a17}       
\end{equation}       
to change the integration variable from $dk_T^2$ to    
$dM^2$.  Then (\ref{eq:a16})      
has the dispersion-like form     
\be     
\sigma_T (\gamma^* p) \; = \; \frac{\alpha}{2\pi} \sum_q    
e_q^2 \int dz      
\frac{dM^2}{(Q^2 + M^2)^2} \: \left \{ M^2 \left [z^2 +    
(1 - z)^2 \right ] + 2 m_q^2      
\right \}.     
\label{eq:b17}     
\ee     
In comparison to (\ref{eq:a12}) we see that    
(\ref{eq:b17}) is a two-dimensional      
integral.  To see the reason for this let us consider massless quarks.     
Then $z = \frac{1}{2} (1 + \cos      
\theta)$ where $\theta$ is the angle between the $q$ and    
the $\gamma^*$ in the      
$q\bar{q}$ rest frame.  The $dz$ integration is implicit    
in (\ref{eq:a12}) as the      
integration over the quark angular distribution in the    
spectral function $\rho$.       
       
We use the additive quark model (AQM) to evaluate    
$\sigma_{q\bar{q} + p}$ in      
(\ref{eq:a16}) which means that each quark is assumed to    
interact with the target      
proton individually.  For forward scattering,    
(\ref{eq:b12}), the momentum of the      
interacting quark is not changed and thus there is no    
interference --- that is the initial      
and final $q\bar{q}$ states are exactly the same.  Now    
the cross section      
$\sigma_T(\gamma^*p)$ of (\ref{eq:a16}) receives    
contributions      
from all       
$M^2$ up to $Q^2$. However to estimate the non-   
perturbative higher twist        
contribution we must note that the additive quark model    
result\footnote{To be        
precise instead of (\ref{eq:a18}) we take the    
$\sigma_{q\bar{q}+p}$ cross sections       
to be $\frac{1}{2}[\sigma(\pi^+p)+\sigma(\pi^-p)]$ for    
the light        
$q\bar{q}$ pairs (or the $\rho$ and $\omega$ VMD    
contributions), and       
to be $\sigma(K^+p)+\sigma(K^-p)-   
\frac{1}{2}[\sigma(\pi^+p)+\sigma(\pi^-p)]$       
for $s\bar{s}$ pairs (or the $\phi$ contribution). We    
take the hadronic       
cross sections from Ref.~\cite{DL}.},       
\begin{equation}       
\sigma_{q\bar{q}+p} \simeq \frac{2}{3} \sigma_{pp},       
\label{eq:a18}     
\end{equation}       
is only valid up to some relatively low mass, $M < M_0$.    
The       
AQM hypothesis may be justified if the separation between    
the $q$ and the       
$\bar{q}, \Delta r \simeq 1/k_T$, is large in comparison    
with the        
$q\bar{q}$ interaction radius $R$ defined by       
\begin{equation}       
\sigma^{\rm inel}_{q\bar{q}} \ = \ \pi R^2 \ \simeq \         
\sigma^{\rm inel}_{pp}/9 \ \simeq 3-5 {\rm mb}.       
\label{eq:a19}     
\end{equation}       
As the separation $\Delta r \simeq 1/k_T$ becomes smaller        
(that is $k_T > 1/R \simeq 0.5~{\rm GeV}$) the quarks    
start to shadow        
each other and        
the cross section $\sigma_{q\bar{q}}$ decreases. As a    
result the        
integration in (\ref{eq:a16}) is effectively restricted    
to the region $M \lapproxeq      
M_0$,       
where $M_0$ may be estimated from (\ref{eq:a17}) using    
$k_T \simeq 0.5$~GeV,      
the constituent quark mass $m_q \simeq 0.35$~GeV, and a    
typical value of        
$z(1-z) \simeq 0.2$. In this way we obtain the upper    
limit        
$M_0^2 \simeq 2~{\rm GeV}^2$.       
We impose this limit on the results that we present    
below.       
       
This part of phase space (that is the domain of small    
$M^2$, small        
$k_T^2$, but $z \sim 0.5$) is not       
that responsible for the leading logarithm behaviour    
generated by DGLAP        
evolution. Rather the $dk_T^2/k^2_T$ behaviour of    
integral comes        
from the alignment        
kinematic configuration of small $z \sim k_T^2/Q^2$ (or    
$1-z \sim k_T^2/Q^2$).       
Then $\bar{Q}^2$ of (\ref{eq:a14}) is of the order of    
$k_T^2$. Hence the        
cross section for       
$q\bar{q}$ scattering on the proton at large $k_T$ (that    
is $k_T^2 \gg       
\Lambda^2_{\rm QCD}$, but $k_T^2 \ll Q^2$) has the QCD    
form        
$\sigma_{q\bar{q}       
+p} \sim 1/k_T^2$, and so (\ref{eq:a16}) becomes        
\begin{equation}       
\sigma_T(\gamma^*p) \sim \frac{\alpha}{Q^2} \int    
\frac{dk_T^2}{k_T^2}.     
\label{eq:a20}       
\end{equation}       
>From (\ref{eq:a17}) we see that this region of phase    
space, $z \sim k_T^2/Q^2$,       
corresponds to $M^2 \sim Q^2$.       
       
Thus, in summary, the higher twist contribution comes    
from the large distance        
domain, $\Delta r \sim 1/k_T$ of small $k_T$ and $M$, but    
\lq\lq large"        
$z \sim 0.5$, where we have to use a non-perturbative    
(AQM) estimate of       
$\sigma_{q\bar{q}+p}$. On the other hand DGLAP arises    
from        
the perturbative small        
distance regime of large $k_T$ and $M \sim Q^2$, but    
small $z$ (or 1--$z$). \\      
       
\noindent {\large \bf Results and Discussion}       
       
Fig.~5 shows the estimates of the higher twist components    
as a fraction of $F_T$       
as a function of $x$ for two values of $Q^2$. The lower    
continuous curve        
corresponds to the effect of gluon rescattering and is    
calculated from (\ref{eq:a10})       
using the MRS(R2) set of partons \cite{MRS}. The upper    
two curves correspond        
to the two alternative ways to calculate the VMD    
contribution. The dotted curve       
corresponds to $\Delta F_T$ calculated from the    
$\rho,\omega$ and $\phi$       
contributions to the standard VMD formula (\ref{eq:a11}),    
while the dashed curve is        
calculated from eq.~(\ref{eq:a16}), using the AQM formula    
(\ref{eq:a18}), with the      
$q\bar{q}$       
invariant mass $M$ restricted to the domain $M^2 < M^2_0    
= 2~{\rm GeV}^2$.       
The two estimates of the VMD component are in good    
agreement with each       
other. We use constituent quark masses ($m_{u,d} = 350$~MeV and $m_s =   
500$~MeV) for the VMD estimates.       
If we were to use current quark masses then $\Delta F_T$    
would be enhanced by        
about $25\%$.       
       
The combined effect of gluon rescattering and the VMD    
contribution is also       
shown (by the heavy continuous curve) on Fig.~5. To    
calculate the total        
effect we use (\ref{eq:a9}) and (\ref{eq:a10}) for the    
rescattering contribution      
together with the       
AQM for the VMD contribution. In the region of $x = {\rm    
few} \times 10^{-4}$       
we see that the two higher twist components are of    
comparable magnitude. However      
the       
energy dependence of the two contributions is quite    
different.       
The \lq perturbative' rescattering       
term involves $(xg(x))^2\propto s^{0.4}$, but actually    
increases       
faster than this due to the $dx^{\prime}/x^{\prime}$    
integration.       
On the other hand       
the \lq non-perturbative' VMD contribution only grows as    
$s^{0.08}$.       
Thus as $x$ increases the VMD component becomes the    
dominant higher       
twist component. The behaviour is evident in Fig.~5. In    
fact at        
$Q^2=4~{\rm GeV}^2$, for example, by $x=0.1$ the higher        
twist contribution (dominated by the VMD component)    
reaches about 17\% of the      
whole value of $F_T$.  On the other hand at very small    
$x$ the main twist-4      
contribution comes from gluon rescattering and at $Q^2 =    
4$~GeV$^2$ and $x =      
10^{-4}$, for example, the total twist-4 exceeds    
$\frac{1}{4}$ of the $F_T$ value.     
     
These are very large higher twist contributions indeed,    
and there is a puzzle why they      
are not evident in fitting to the data. One possibility    
is that in fitting the data the input       
distribution together with the twist-2 DGLAP evolution is    
able in a limited       
kinematical domain, say $Q^2\sim 3 - 10~{\rm GeV}^2$, to    
mimic the contribution       
coming from higher twists. If this is the case, then we    
should subtract       
the higher twist (or at least the VMD contribution which    
is more or less known)       
from the DIS data before we perform the DGLAP analysis.    
In other words in the       
fit to the data we should include the  VMD contribution    
(and maybe also       
gluon screening) as well as the conventional DGLAP       
contributions.  An alternative possible resolution of the    
puzzle is that there exists yet       
another important source of higher twist which enters    
with a negative sign       
such that it partially cancels out the higher twist component.  For    
instance renormalon models      
have been used to estimate power corrections to deep    
inelastic scattering      
\cite{SMMS}.     
       
We conclude that there are theoretical reasons to expect    
very        
significant higher twist        
contributions to $F_2$, even for $Q^2$ as high as    
$10~{\rm GeV}^2$. Surprisingly       
there is almost no evidence for them in the    
data\footnote{In the latest global analysis      
\cite{MRST} partons have been obtained by fitting to data    
with $Q^2 > 2$~GeV$^2$,      
and then the analysis was repeated with the omission of    
data below $Q^2 =      
10$~GeV$^2$.  The gluon obtained at $Q^2 = 10$~GeV$^2$    
from the latter analysis      
is some 5\% higher at small $x$ than the gluon from the    
former analysis which      
contained low $Q^2$ data.  This could be the first    
indication that some non-twist-2      
contribution is hidden in the data.}. We have been able    
to estimate the        
size of the two major higher twist contributions coming,    
first, from vector       
meson dominance and, second, from gluon rescattering in    
which the four-gluon       
state is composed of two colour singlets. These two    
higher twist contributions        
have very different energy dependences and there is an    
increasingly        
significant resultant higher twist effect as $x$    
decreases due to the       
increasing importance of the gluon rescattering term. Of    
course it is possible that      
$F_2$ may be subject to other higher twist contributions    
which partially cancel the      
effects of the two components that we have estimated and    
which are expected to dominate.  This exploratory       
study makes clear that, on the one hand, the higher twist    
contribution        
needs more detailed theoretical analysis while, on the    
other hand, it is       
important to extract as much experimental information as    
possible by fitting       
the $Q^2$ dependence of precise data on $F_2(x,Q^2)$ in    
the region of       
$Q^2 \lapproxeq 10~{\rm GeV}^2$.  \\     
       
\noindent {\large \bf Acknowledgements}       
       
MGR thanks the Royal Society, INTAS (95-311) and the    
Russian Fund of       
Fundamental Research (98-02-17629), for support.

\newpage       
       
\noindent {\large \bf Figure Captions.}       
       
\begin{itemize}       
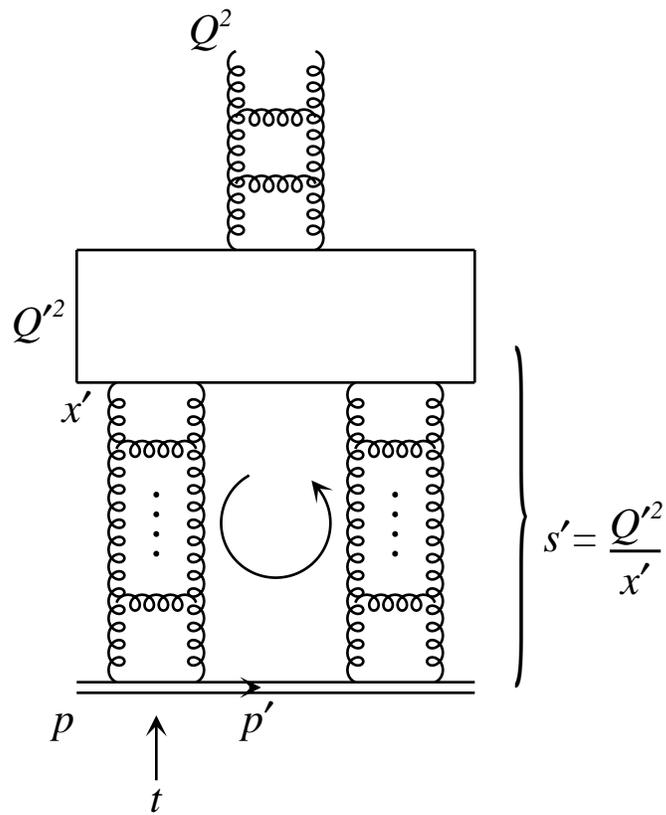
\item [Fig.~1] An absorptive or gluon rescattering diagram giving rise to a twist-4   
contribution.       
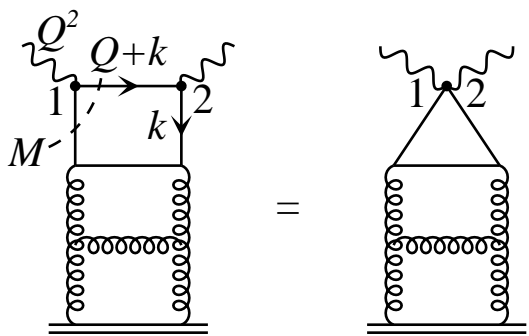
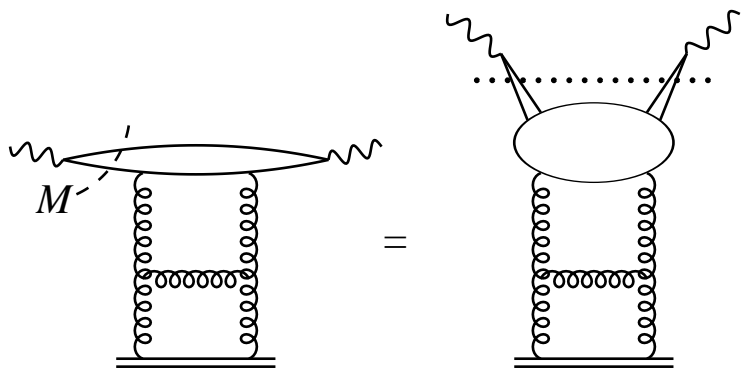
\item [Fig.~2] An illustration of the space-time structure of a conventional twist-2   
DGLAP contribution and a twist-4 vector meson dominance (VMD) contribution. In   
the latter case the dotted lines makes the 4 quark structure evident.       
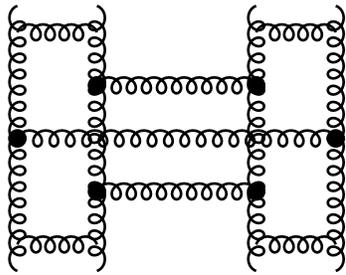
\item [Fig.~3] An example of an interaction between the colour singlet two-gluon   
ladders which gives rise to the factor $K$ in (\ref{eq:a7}).       
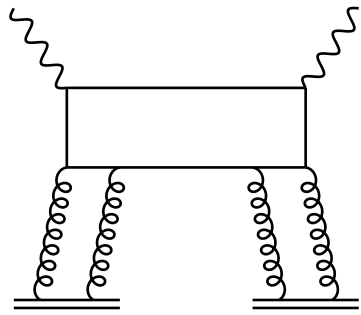
\item [Fig.~4] A contribution to twist-4, $F_T^{(4)}$, associated with the quark,   
rather than the gluon, distribution.       
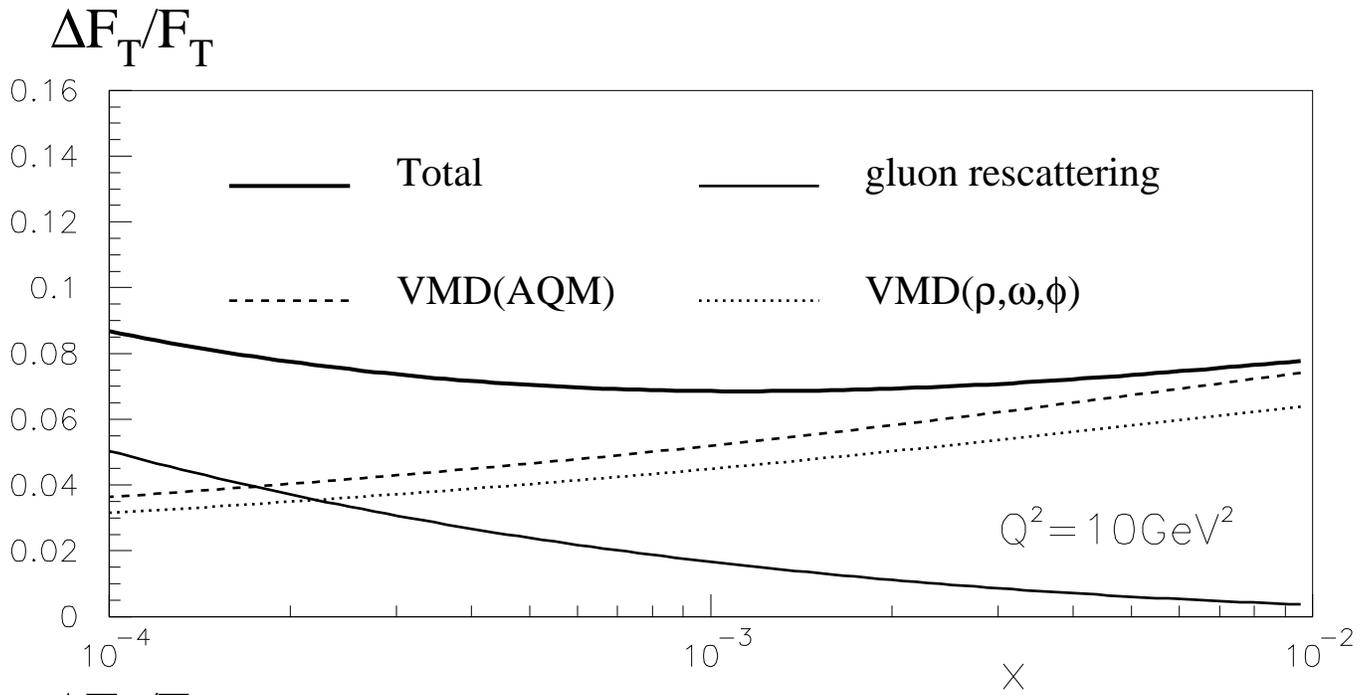
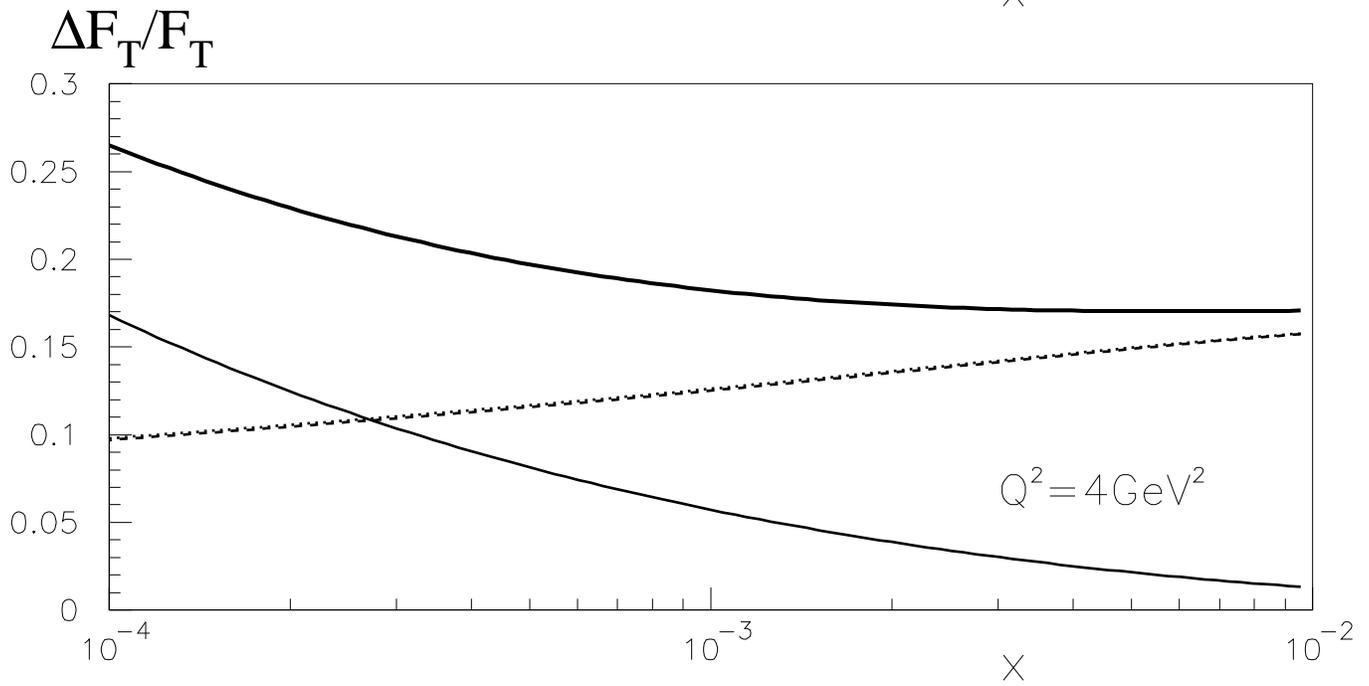
\item [Fig.~5] The higher twist, $\Delta F_T$, contributions compared to $F_T$ at   
$Q^2 = 10$ and $4~{\rm GeV}^2$. Two estimates of the VMD fraction are shown;   
they are essentially coincident at $Q^2 = 4~{\rm GeV}^2$. The fraction obtained   
from the sum of the gluon rescattering and the VMD(AQM) higher twist contributions   
is shown as the heavy continuous curve.  
  
\end{itemize}       
       
\end{document}